\documentclass[letter]{aa} 
\usepackage{graphicx}
\usepackage{txfonts}
\usepackage{natbib}
\bibpunct{(}{)}{;}{a}{}{,} 

\begin{document}

\title{Observations of a rotating macrospicule associated with an X-ray jet}

\author{S. Kamio\inst{1} \and W. Curdt\inst{1} \and L. Teriaca\inst{1} \and B. Inhester\inst{1} \and S.K. Solanki\inst{1,2}}
\institute{Max-Planck-Institut f\"ur Sonnensystemforschung (MPS),
Max-Planck-Str. 2, 37191 Katlenburg-Lindau, Germany\\
\email{skamio@spd.aas.org}
\and
School of Space Research, Kyung Hee University, Yongin, Gyeonggi 446-701, Korea
}

    \date{Received ; accepted }

\abstract{}
{We attempt to understand the driving mechanism of a macrospicule
and its relationship with a coronal jet.}
{We study the dynamics of a macrospicule and an associated coronal jet
captured by multi-spacecraft observations.
Doppler velocities both in the macrospicule and the coronal jet
are determined by EIS and SUMER spectra.
Their temporal evolution is studied using X-ray and \ion{He}{ii} $\lambda$304 images.}
{A blueshift of $-120\pm15$~km~s$^{-1}$ is detected
on one side of the macrospicule,
while a redshift of $50\pm6$~km~s$^{-1}$ is found at the base of
the other side.
The inclination angle of the macrospicule inferred from a stereoscopic
analysis with STEREO
suggests that the measured Doppler velocities can be attributed to a rotating
motion of the macrospicule rather than a radial flow or an expansion.}
{The macrospicule is driven by the unfolding motion of a twisted magnetic flux rope, while the associated X-ray jet is a radial outflow.}


\maketitle

\section{Introduction}

Macrospicules are spiky features extending from the limb,
which were first observed in \ion{He}{ii} $\lambda$304 images
\citep{bohlin1975}.
They are typically 5\arcsec~ to 50\arcsec~ long
and 5\arcsec~ to 30\arcsec~ wide and have a lifetime
of 5 to 40 minutes.
Macrospicules are formed at temperatures lower than
$3 \times 10^5$~K
and are not seen in emission at higher temperatures.
Since they occurt primarily within coronal holes,
it has been suggested that open magnetic field lines
might play a role in macrospicule formation.
\citet{moore1977} compared images in
\ion{He}{ii} $\lambda$304 and H$\alpha$
and concluded that H$\alpha$ macrospicules are the
chromospheric counterpart of EUV macrospicules.
They showed that flares in X-ray bright points coincided
with H$\alpha$ macrospicules.
However, the long exposure time needed for X-ray observations
did not allow the temporal evolution to be traced.
\citet{habbal1991} reported microwave observations of
macrospicules in polar coronal holes.
They found pinching-off in the upper part of some macrospicules
and the separation of plasmoids.
They suggested that macrospicules could supply
mass to the solar wind.

\citet{pike1998} reported blueshifted and redshifted
emission on opposite sides of macrospicules and
interpreted them as rotating motion.
However, the problem of spectroscopic observations
is that a sequence of spectra is a mixture of
spatial structure and temporal evolution
because scanning times are often comparable
to the lifetime of macrospicules.
Considerable morphological evidence of
helical signatures in jets has been found
\citep{shimojo1996,patsourakos2008},
although the velocity of the helical motion could not
be determined.

Apart from macrospicules, it has been reported
that numerous X-ray jets occur in polar coronal holes
\citep{cirtain2007, savcheva2007, shimojo2007}.
Their frequent occurrence and the associated high speed outflows
of up to 800~km~s$^{-1}$ suggest that they may play a role
in solar wind acceleration.
Many observations detect coronal jets associated with
cool features.
\citet{canfield1996} reported the simultaneous occurrence of
H$\alpha$ surges and X-ray jets.
\citet{kim2007} found a correlation of X-ray jets with cooler
material showing strong blue shift on the disk.
\citet{raouafi2008} discovered plume haze rising from locations
of polar coronal jets.

Our new observations detected the detailed structure and
evolution of a macrospicule associated with a coronal jet.
The combination of
the EUV Imaging Spectrometer \citep[EIS;][]{culhane2007}
on {\it Hinode} \citep{kosugi2007} and
Solar Ultraviolet Measurements of Emitted Radiation instrument \citep[SUMER;][]{wilhelm1995} on the Solar and Heliospheric Observatory \citep[SOHO;][]{domingo1995} allowed us to
measure the line of sight (LOS) motions of both macrospicule and
coronal jets.
At the same time, the X-Ray Telescope \citep[XRT;][]{golub2007} on {\it Hinode}
and SECCHI \citep{howard2008,wuelser2004}
on {\it STEREO} \citep{kaiser2008} traced the
evolution of the coronal jet and the macrospicule.
In addition, the 3D geometry of the macrospicule
was estimated from a stereoscopic analysis of SECCHI image
pairs.

\section{Observations and data reduction}

The northern polar coronal hole was observed by {\it Hinode}
and SUMER on 4 November 2007.
In the coordinated observations,
70\arcsec~ wide scans were repeatedly carried out by EIS and SUMER
at 60 minute cadence.
This movie mode was intended to increase the probability of
capturing an identical event.
Strong emission in
\ion{Fe}{xii} $\lambda$195.12 ($T_{\mbox{\small e}} = 1.2 \times 10^6$K) and
\ion{He}{ii} $\lambda$256.32 ($T_{\mbox{\small e}} = 5.0 \times 10^4$K) from EIS and
\ion{Ne}{viii} $\lambda$770.41 ($T_{\mbox{\small e}} = 6.3 \times 10^5$K) and
\ion{O}{iv} $\lambda$790.20 ($T_{\mbox{\small e}} = 1.6 \times 10^5$K)
from SUMER were selected and analyzed.
The exposure time of 45 sec was sufficient to collect
enough counts and match the telemetry constraints.
EIS and SUMER data were calibrated using
standard procedures provided in the Solar Software tree.
The spectra were fitted by a Gaussian to derive
radiance and Doppler shifts of emission lines.

The $1\sigma$ value of the Gaussian fit is assumed to measure
the uncertainty in the Doppler shift measurements.
Velocities are inferred by the assuming that the mean velocity
in the quiet region is zero.
Co-alignment of EIS and SUMER was checked by identifying
bright points in the coronal hole that were
stable during the observation.
The estimated uncertainty in the alignment is $\approx 1$\arcsec,
the pixel size of EIS and SUMER.

XRT took X-ray images with
Al\_poly and Ti\_poly filters,
which are sensitive to low temperature coronal emission
down to $1\times10^6$ K.
The time series with a 1 minute cadence {\bf allowed
us to study the detailed evolution.}

SECCHI/EUVI onboard the two {\it STEREO} spacecraft
recorded \ion{He}{ii} $\lambda$304 broadband filter images
obtained with a 10 minute cadence and 3.2\arcsec~ resolution.
The separation angle of $39^{\circ}$ between {\it STEREO-A}
and {\it B} allowed us to derive the 3D structure of the macrospicule.
For the stereoscopic projection of the \ion{He}{ii} emission from the
macrospicule, we separated the macrospicule emission from the image background.
The macrospicule images were then divided into 20 equidistant slices
of stacked epipolar planes and for each plane the 3D centroid of the observed
intensity was constructed.
Only the principal axis of the macrospicule was determined since
fine structure inside the macrospicule was not resolved in the images.
For a general description of the stereoscopic
reconstruction, we refer to \cite{inhester2006}.
We ignore the slight rotation of the Sun relative to this coordinate system
during the 20 min of the {\it STEREO} observations.
To obtain accurate results, we had to correct the Sun centre position
given in the file headers by a fraction of a pixel (W\"ulser, private communication).
At the time of these observations, these corrections amounted to
$\Delta x_A=0.0$ pixel, $\Delta y_A=-0.7$ pixel, $\Delta x_B=0.2$ pixel, and $\Delta
y_B=0.8$ pixel.

\section{Results}

\begin{figure}
\centering
\includegraphics[width=8cm]{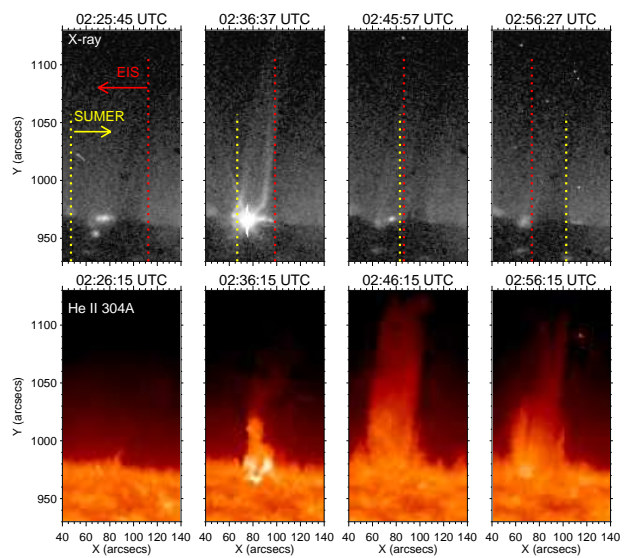}
\caption{{\it Top row:} Sequence of XRT Al\_poly filter images
showing an X-ray jet.
Red and yellow lines respectively indicate EIS and SUMER slit locations
at corresponding times. Arrows show scanning directions of EIS and SUMER.
Horizontal and vertical axes indicate heliocentric coordinates.
{\it Bottom row:} {\it STEREO}-A EUVI \ion{He}{ii} $\lambda$304 images. 
\label{fig_xrt_euvi}}
\end{figure}

The sequence of XRT images indicate that five coronal jets
occurred at the same location close to the limb between
18:00 UTC on 3 November and 03:00 UTC on the
following day.
EIS and SUMER successfully captured
the most significant event hereof.
Figure \ref{fig_xrt_euvi} shows
time series of X-ray and \ion{He}{ii} $\lambda$304 images.
Dashed lines plotted on XRT images
show the slit locations of EIS and SUMER
at the corresponding times.
The coronal jet consisted of two threads that originated in an
X-ray bright point and extended into the upper corona.
The front of the jet reached $8\times10^4$ km above the bright point
when it attained maximum brightness at 02:36 UTC.
The apparent propagation velocity projected on to the plane of the sky
is determined by measuring the front height of the jet.
The apparent upward velocity inferred from XRT images
between 02:31 UTC and 02:36 UTC is $320\pm 50$~km~s$^{-1}$.

In \ion{He}{ii} $\lambda$304 images at 02:36 UTC,
the macrospicule appeared with bright points at the footpoint.
The axis of macrospicule deviates by 50 arcsec from the ecliptic pole.
If viewed from the direction of the Earth,
the estimated displacement is 3 arcsec,
which was not compensated for in Fig. \ref{fig_xrt_euvi}.
EUVI images in Fig. \ref{fig_xrt_euvi} are rotated so that the
solar North is up.
The axis of the macrospicule is located
between the two threads of the X-ray jet.
The X-ray jet disappeared by 02:45 UTC, but the macrospicule
grew in height and width.
The macrospicule attained its maximum height of $1\times 10^5$~km
at 02:46 UTC, when it expanded to $2.6\times 10^4$~km in width.
The 10-minute cadence of \ion{He}{ii} images
is insufficiently short to study the evolution of
the macrospicule, but it provides a
velocity averaged over 10 minutes.
The upward and horizontal propagating velocities
between 02:36 UTC and 02:46 UTC
are $130\pm 30$~km~s$^{-1}$ and $25\pm 5$~km~s$^{-1}$, respectively.
Its upper part became fainter at 02:56 UTC,
while the radiance slightly increased at the bottom.
The macrospicule had disappeared in the following
exposure at 03:06 UTC.

\begin{figure}
\centering
\includegraphics[width=8cm]{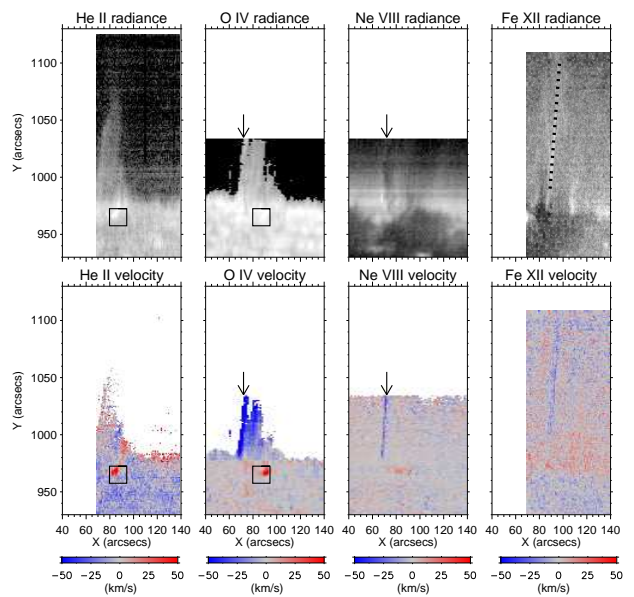}
\caption{Radiance and Doppler velocity maps in
\ion{He}{ii}, \ion{O}{iv}, \ion{Ne}{viii}, and \ion{Fe}{xii}.
Boxes in the \ion{He}{ii} and \ion{O}{iv} panels
mark the location of the redshifted point.
Arrows indicate the location of the blueshifted feature
observed by SUMER.
Dotted line on the \ion{Fe}{xii} image marks the locations
of the bright
streak on the right side of the macrospicule.
\label{fig_vel}}
\end{figure}

Radiance and Doppler velocities measured by EIS and SUMER
are displayed in Fig. \ref{fig_vel}.
The EIS and SUMER slits targeted the jet from opposite
directions and successfully captured both edges of
the jet at the same time.
This is an extremely fortunate case, in which both spectrometers had their
slits at the right position and could provide
spectroscopic information about a transient, short-lived event.

The SUMER slit scanned from the left side
and enhanced emission in \ion{Ne}{viii} was detected
at 02:38 UTC,
which is marked in Fig. \ref{fig_vel} by arrows
in the lower \ion{Ne}{viii} panel.
This brightening lies at the foot of a blueshifted streak of about
 $-25\pm5$~km~s$^{-1}$ observed in \ion{Ne}{viii},
which corresponds to the left thread of the X-ray jet
in the second column of Fig. \ref{fig_xrt_euvi}.

The excess emission and the blueshift in \ion{Ne}{viii}
quickly disappeared as SUMER scanned the inner part of the macrospicule,
where the \ion{Ne}{viii} emission was even weaker than the background
corona, probably due to absorption by cool dense plasma.
Coexisting enhanced emission in \ion{O}{iv}
indicates the existence of cool material.
\citet{anzer2005} showed that dark features in
EUV coronal emission lines are caused by the presence
of cool material in the corona.

At the same time as the blueshifted streak was observed in \ion{Ne}{viii},
offlimb \ion{O}{iv} emission appeared, which is unusual for that line.
It extended to at least $4.4 \times 10^4$~km above
the limb, which was at the edge of SUMER's FOV.
The \ion{O}{iv} emission exhibited a
blueshift ranging from $0\pm10$~km~s$^{-1}$ at the limb to
 $-120\pm15$~km~s$^{-1}$
at the top of the slit.
As SUMER scanned across to the right,
the blueshift gradually decreased
and almost vanished at the right edge of the macrospicule.

The EIS slit approached the jet from the right side
and encountered enhanced coronal emission in \ion{Fe}{xii}
at 02:40 UTC, which is marked by a dotted line at $x = 90$\arcsec.
The blueshifted streak in \ion{Fe}{xii} 
corresponds to the right thread of the X-ray jet
in the second column of Fig. \ref{fig_xrt_euvi}.
The enhanced emission is blueshifted by $-20\pm8$~km~s$^{-1}$, almost
the same velocity as the \ion{Ne}{viii} at $x = 70$\arcsec~ 
on the other side of the structure.
On the left side of the enhanced emission, \ion{Fe}{xii} radiance
became weaker than the surrounding corona, probably because of
absorption by cool material,
similar to the darkening in \ion{Ne}{viii}.

In the \ion{He}{ii} velocity, we only studied the ondisk region
because the velocity of the offlimb region is largely affected by coronal
emission in the red wing of
\ion{He}{ii} $\lambda$ 256.32, namely by
\ion{Si}{x} $\lambda$ 256.37,
\ion{Fe}{xii} $\lambda$ 256.41, and
\ion{Fe}{xiii} $\lambda$ 256.42
\citep{young2007}.
Since the \ion{He}{ii} emission has small a contribution in the offlimb
region, increase or decrease in coronal emissions causes an artificial
spectral shift.
A significant brightening and redshift of $+50\pm6$~km~s$^{-1}$
was found in \ion{He}{ii} at 02:47 UTC,
which is marked by a box in the first column of Fig. \ref{fig_vel}.
Since coronal emission did noticeably change there,
the observed redshift can be attributed to
\ion{He}{ii} rather than to blending coronal emission.
The SUMER slit reached the region later and
also detected a $+50\pm4$~km~s$^{-1}$ redshifted point in \ion{O}{iv}
at 02:49 UTC, which was located $5\times 10^3$ km
to the right of the redshift in \ion{He}{ii}
(box in the second column of Fig. \ref{fig_vel}).
The distance of $5\times 10^3$ km is greater than the uncertainty
in co-alignment, hence it is thought to be a drifting redshifted
feature caused by the expansion of the macrospicule seen in Fig.
\ref{fig_xrt_euvi}.
\citet{canfield1996} studied several H$\alpha$ surges adjacent
to X-ray jets and reported
whiplike motion of filaments and moving blueshifted features
in H$\alpha$ surges.
The drifting redshifted feature in the present paper
could be a similar phenomena.

\begin{table}
\caption{Parameters of the macrospicule's principal axis.
  The footpoint position is given in HEEQ spherical coordinates
  The principal direction $\Delta \theta$ and $\Delta \phi $
  show the deviations from the radial direction.
  In the last column, the angle of the principal axis relative to
  the LOS from the Earth is listed.}
\label{stereo_table}
\centering
\begin{tabular}{cccccc}  \hline \hline
Time    & \multicolumn{2}{c}{footpoint in HEEQ}
        & \multicolumn{2}{c}{principal direction}
        & angle to Earth \\
(UTC)   & lon $\phi$  &  lat $\theta$
        & $\Delta \phi$ & $\Delta \theta$
        & \\  \hline
02:36  & 111.59 &  84.84   &  2.60 &  -3.63  &  84.36         \\
02:46  & 113.69 &  85.44   &  4.73 &   2.73  &  84.13         \\
02:56  & 110.05 &  85.63   &  2.49 &   7.89  &  92.37         \\
 \hline
\end{tabular}
\end{table}

\begin{figure}
\centering
\includegraphics[height=6.5cm]{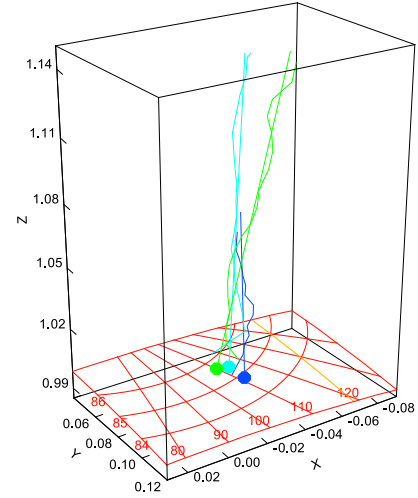}
\caption{Oblique view of the macrospicule reconstructed from
{\it STEREO} SECCHI images.
The centroid curve, fitted principal line, and footpoints are colour
coded for the three times of observation: 02:36 UTC (dark blue),
02:46 UTC (light blue), 02:56 UTC (green).
The Cartesian HEEQ coordinates are given in units of the solar radius
along the outer square box.
The solar surface is covered with the corresponding spherical grid in red.
Earth is in the positive x direction (lower left).
The apparent limb from Earth is indicated by the orange curve on the
solar surface.
\label{fig_stereo}}
\end{figure}

The reconstructions of the 3D orientation of the macrospicule
were performed in the Heliocentric Earth Equatorial (HEEQ)
coordinate system, whose Z-axis is along
the Sun's rotation axis and for which zero longitude is the meridian of Earth.
For each time, the reconstructed 3D centroid curve was fitted to
a principal line. The intersection of this line with the solar
surface defined the nominal foot point position of the macrospicule. 
Table~\ref{stereo_table} lists the main parameters of the reconstructed
macrospicule principal axes at the three {\it STEREO} observation times.
In Fig.~\ref{fig_stereo}, we display these results graphically,
color-coded for the different observation times.
The location of the macrospicule is very close to but clearly on the
earth-ward side of the apparent limb as seen from Earth.
Small values of $\Delta \phi$ and $\Delta \theta$ indicate that
the macrospicule axis is close to the radial direction.
The last column in Table~\ref{stereo_table} lists the angle that the
reconstructed macrospicule axis makes with the view from Earth, or 
identically from SUMER and EIS. Since this angle deviates by
no more that 6 degrees from 90 degrees, the macrospicule axis
is almost perpendicular to the LOS.

\section{Discussion and summary}

Multi-spacecraft observations have monitored the detailed
evolution of a macrospicule associated with a coronal jet.
The macrospicule exhibited a significant blueshift on the left side
and a drifting redshifted feature on the right side,
while the coronal jet showed blueshifts on both sides.
The 3D reconstruction of the macrospicule
has allowed us to distinguish between radial flow and helical motion.
Assuming an apparent propagation speed of $130\pm 30$~km~s$^{-1}$
corresponding to a radial outflow in the macrospicule,
this could account for a LOS velocity of $-15$~km~s$^{-1}$,
which is far smaller than $-120\pm15$~km~s$^{-1}$
observed on the left edge of the macrospicule.
Similarly, the apparent horizontal expansion of the macrospicule
could only produce a LOS velocity of $-25\pm 5$~km~s$^{-1}$.
Hence the Doppler shifts observed by SUMER can hardly be
attributed to a radial outflow or an expansion of the macrospicule.
Oppositely directed flows on either side of the macrospicule
are indicative of a rotating motion along its axis.
The coronal jet exhibited smaller Doppler shifts.
If we assume the same inclination angle as for the macrospicule,
a radial outflow of $320\pm 50$~km~s$^{-1}$, as deduced from the
expansion
of the jet in a series of X-ray images, could account for
a LOS velocity of $-30$~km~s$^{-1}$, which is comparable to
the detected Doppler shifts of
$-25\pm5$~km~s$^{-1}$ in \ion{Ne}{viii} and $-20\pm8$~km~s$^{-1}$ in \ion{Fe}{xii}.
The results suggest that the coronal jet and the macrospicule
have different structures;
the coronal jet consists of an outflowing hot plasma,
while the macrospicule is formed by a cool plasma with rotating motion.
The different evolutions of the X-ray jet and
the macrospicule also supports this scenario (Fig. \ref{fig_xrt_euvi}).
The two-threaded X-ray jet travelled at an apparent speed of
$320\pm 50$~km~s$^{-1}$, while the macrospicule grew at
$130\pm 30$~km~s$^{-1}$.
\citet{nistico2009} classified this event as helical
by studying a sequence of {\it STEREO}/SECCHI images.
The \ion{He}{ii} $\lambda$304 image at 02:36 UTC
(the second column of Fig. \ref{fig_xrt_euvi}) shows
that the macrospicule has a peculiar shape, which might be interpreted
as a twisted flux rope.
\ion{Fe}{ix/x} $\lambda$171 images show a pre-existing plume
above the jet site.
The plume might be connected to the jet, although no noticeable change
was found in the plume structure before and after the jet
(not shown in the present paper).

\citet{yokoyama1995}
performed two-dimensional MHD simulations
and demonstrated that magnetic reconnection
between emerging flux and open fields can produce
X-ray jets and nearly co-located H$\alpha$ surges.
Both hot and cool plasma are expected to be
accelerated more or less at the same time.
X-ray jets and neighboring H$\alpha$ surges
were simultaneously observed in active regions
\citep{schmieder1995}.
\citet{canfield1996} observed helical motions of H$\alpha$ surges
and converging magnetic footpoints and interpreted them in the context of
\citet{shibata1986} model in which
the successive reconnections release the magnetic twist into
open magnetic fields and the twist drives the surge.
The same mechanism can also work for the macrospicule.
Therefore, it is likely that magnetic reconnection
initiated the coronal jet and the macrospicule
at the same time.

\citet{pariat2009} carried out
three dimensional MHD simulations in which
twist is imposed on axisymmetric magnetic fields.
They demonstrated that reconnection
between twisted magnetic fields and open
fields produces torsional Alfv\'en waves
that drives plasma upflow with helical motion.
Here we speculate that magnetic reconnection produced
the X-ray jet and the relaxation of
helical magnetic fields resulted in the macrospicule.
\citet{tian2008} found that hot and cool loops in
a bright point showed different orientations
and interpreted them as a twisted loop system.
If twisted magnetic loop systems are commonly
related to bright points or jets,
they could provide the original configuration leading to
rotating motion in macrospicules.
However, there are still issues to be solved in jet models.
The origin of the twist is not fully understood.
\citet{pariat2009} assumed that an axisymmetric field configuration
builds up enough energy to produce the helical structure,
which may not be realistic.

It seems as if most of the macrospicule
falls back to the surface in the end.
In the decay phase of the macrospicule,
\ion{He}{ii} indicates enhancement at the bottom
and fading out at the upper part
(Fig. \ref{fig_xrt_euvi}).
The fading of the upper part could be
caused by heating of cool plasma, but no clear signature of
heating was found in hotter emission lines.
The radiance enhancement at the bottom can be
interpreted as a density increase caused by falling
back cool plasma.
\citet{culhane2007a} analyzed slot images
obtained with EIS and found post-jet enhancements
in cool emission lines, which were also interpreted as
falling back jet plasma.
Although they did not identify macrospicules as the cause,
the delayed enhancement in cool lines including
\ion{He}{ii} that they found could be due to macrospicular material.
However, part of the material might be ejected into the corona,
because blobs pinching off from macrospicules were reported
\citep{habbal1991}.

In our study, both 3D orientation and velocities of a
macrospicule have been determined for the first time.
These results clearly detect a rotating motion in
the macrospicule and show only radial outflow
in the coronal jet.
Macrospicule and jet models in 3D need to reproduce
these properties.
\citet{nistico2009} carried out a statistical
study of coronal jets seen by {\it STEREO}/SECCHI
and reported many events associated with
\ion{He}{ii} $\lambda$304 features, including this event.
This proves that this event, in which the coronal jet and
the macrospicule are initiated at the same time, is
a common phenomenon in the solar atmosphere.

\acknowledgements
{\it Hinode} is a Japanese mission developed and launched by ISAS/JAXA,
with NAOJ as domestic partner and NASA and STFC (UK) as international
partners.
It is operated by these agencies in co-operation with ESA and NSC (Norway).
The SUMER project is financially supported by DLR, CNES, NASA, and the ESA
PRODEX Programme (Swiss contribution).
SUMER is part of {\it SOHO} of ESA and NASA.
{\it STEREO} is a project of NASA. The data from the SECCHI instrument
used here was produced by an international consortion led by the
Naval Research Lab (USA).
The STEREO contributions by the MPS were supported by DLR grant 50OC0501.
This work has been partially supported by WCU grant No. R31-10016 funded
by the Korean Ministry of Education, Science, and Technology.

\bibliographystyle{aa}
\bibliography{reference.bib}

\end{document}